# Single neutral substitutional 3d-transition metal in GeTe and GeSb$_2$Te$_4$ by the screened exchange functional

H. Li[*], L. P. Shi

Department of Precision Instrument, Centre for Brain Inspired Computing Research, Tsinghua University, Beijing, 100084

[*]*Corresponding author email: li_huanglong@mail.tsinghua.edu.cn*

Abstract

Fe doped GST has shown experimentally the ability to alter its magnetic properties by phase change. In this work, we use screened exchange hybrid functional to study the single neutral substitutional 3d transition metal (TM) in crystalline GeTe and GeSb$_2$Te$_4$. By curing the problem of local density functional (LDA) such as over delocalization of the 3d states, we find that Fe on Ge/Sb site has its majority d states fully occupied while its minority d states are empty, which is different than previous predicted electronic configuration by LDA. From early transition metal Cr to heavier Ni, the majority 3d states are gradually populated until fully occupied and then the minority 3d states begin to be filled. In order to study the magnetic contrast, we use lower symmetry crystalline GeTe and GeSb$_2$Te$_4$ as the amorphous phases, respectively, which has been proposed to model the medium range disordering. We find that only Co substitution in r-GeSb$_2$Te$_4$ and s-GeSb$_2$Te$_4$ shows magnetic contrast. The experimental magnetic contrast for Fe doped GST may be due to additional TM-TM interaction, which is not included in our model. It can also be possible that these lower symmetry crystalline models are not sufficient to characterize the magnetic properties of real 3d TM doped amorphous GST.

Introduction:

Optical, electrical and magnetic storage are three main data storage technologies and one technology can outperform the other under different specifications. Each of them often requires unique storage medium. One exception is the wide application of phase change materials (PCMs) like Ge$_x$Sb$_y$Te$_z$ (GST) in both the optical memory [1,2] and random access memory [3-5]. PCMs are well known for fast and reversible switching of both optical and electrical properties by phase change between their crystalline and amorphous phases [6]. However, until now there is no reported magnetic device using PCMs. It is still tempting to find one material which shows all controllable optical, electrical and magnetic properties and preferably allows ready integration with the existing technologies. GST is nonmagnetic but magnetic properties can be engineered into GST by doping with Fe [7,8]. It is highly desirable to be able to control the magnetism once the magnetic medium has been prepared and put into use. External electrical field [9 and ref. therein] and spin-polarized current [10,11] have shown the ability to control the magnetism without changing the atomic bonding network. Charge carrier redistribution and spin torque induced spin flipping are responsible for the magnetic switching, respectively. The Fe-GST, on the other hand, demonstrate experimentally the magnetic contrast by phase change [7,8], which may suggest an alternative route of fast manipulation of the magnetism by phase change. In this work, we focus on two typical PCMs, namely GeTe and GeSb$_2$Te$_4$.

Transition metal (TM) doped magnetic GeTe has first been studied by Rodot et al [12] even before the reversible resistance switching by phase change was found [13]. TM-GeTe has been being studied as a standard diluted magnetic semiconductor (DMS). Its magnetic properties and phase change related properties have been considered separately. Until 2008, Song et al [7] first demonstrated the phase change driven magnetic switching in Fe-$Ge_2Sb_2Te_5$ as both a DMS and a PCM. Tong et al [8] found similar effects in GeTe. Theoretically, the magnetism in TM-GST has been studied by first principle calculation [14-19]. The exchange-correlation effect of the electrons was presented by the local density approximation (LDA) or the generalized gradient approximation (GGA) there. However, it is well known that the LDA and GGA underestimate the band gap in semiconductors and insulators. The underestimation of the band gap is often manifested by a too low conduction band minimum which may spuriously place the impurity orbital into the host conduction band. The electrons will then drop into the conduction minimum, resulting in false occupancy and incorrect magnetism [20]. LDA/GGA+U has also been used and qualitatively similar results to those of LDA/GGA were reported [15]. However, LDA/GGA+U does not solve the problem since it shifts the unoccupied 3d impurity level to yet higher energies [20]. LDA and GGA also do not describe well localized states such as in the 3d-TM incorporated systems where the TM 3d states are described as too shallow, resulting in a large hybridization with the host p states. Consequently, the TM 3d states are over-delocalized [21]. GGA+U localizes the 3d states but leads to other problems [21]. Hybrid functionals such as the screened functional (sX) are useful here to better describe the electronic structures of TM-GST. In this work, by calculating the atomic density of states of TM-GST we try to explain the magnetism by the sX.

As stated above, the promising perspective of TM-GST is the magnetism modification by phase change. The remarkably large optical dielectric constant contrast and electrical resistance contrast of GST have led to extensive studies of the nature of atomic bond and local structure. The uniqueness of the crystalline phase is its resonant bonding which requires a longer-range order than the conventional electron pair bond of the 8-N rule [22]. The crystalline GeTe shows a rhombohedral (r-) structure [23]. The crystalline $GeSb_2Te_4$ shows a rocksalt (r-) structure with Te atoms on one sublattice and Ge as well as Sb atoms and 25% vacancies on the other [24]. The difference in bonding of the crystalline and amorphous phases is now understood by the local order [25-28]. This can be modeled in terms of either the nearest-neighbor coordination number [25,26] or the orbital alignment irrespective of the coordination number [27]. In the first model, Welnic et al [26] proposed a lower symmetry crystalline phase spinel (s-) structure for the amorphous $GeSb_2Te_4$, in which Ge atoms occupy tetrahedral and Sb and Te atoms octahedral positions. The Ge coordination number changes from six-fold in the crystalline phase to four-fold in the amorphous phase. In the second model, Huang et al [27] proposed the orthorhombic (o-) GeTe for the amorphous GeTe. It retains the 3-3 bonding of r-GeTe but loses the p orbital alignment. In this work we use lower symmetry crystalline phase o-GeTe and s-$GeSb_2Sb_4$ as the representation of the corresponding amorphous phases, although direct modelling of the amorphous phase by the melt-and-quench molecular dynamic (MD) method exists, characterizing more complex ring patterns in the structure [29,30]. We want to see the magnetic properties of various single TM interstitial in these interpretative amorphous models and find out whether they act as local magnetic moment-carrying center for possible collective ferromagnetic interaction to occur.

Computational methods

This work is carried out by first principle calculation in the Cambridge Serial Total Energy Package (CASTEP) [31,32]. The calculation of single TM interstitial is studied based on the supercell method. The lattice constants and the internal atomic structures of r-GeTe, o-GeTe, r-GeSb$_2$Te$_4$ and s-GeSb$_2$Te$_4$, are relaxed by the GGA functional [33] with 380 eV cutoff energy of the plane wave basis set and ultrasoft pseudo-potentials. The lattice constants are then fixed and one 3d TM atom (Cr, Mn, Fe, Co, Ni) substitutes one of the Ge or Sb atoms. The point defect geometries are further relaxed by the GGA functional. Based on the calculated ranges of magnetic interaction in other point defect systems [34], we assume that the TM-TM magnetic interaction is negligible in our supercell model. We then apply sX-LDA [35] functional to the GGA relaxed atomic structures to study the electronic structures. In sX-LDA calculation, 750 eV cutoff energy and norm conserving pseudopotentails are used. Considering the lattice constants of our supercell model, Gamma point sampling is used. The total energy of the supercell is converged until below $1\times10^{-5}$ eV. Throughout the calculation, the spin states of all the atoms are allowed to relax in order to reach the energetically preferred state of the system. We want to point out that the spin calculation performed here is collinear, assuming the magnetism can be characterized by atomic moments all aligned to some global quantization axis. Non-collinear spin calculation for probable more complex magnetism is subject to further study.

Results and discussion

Bulk r-GeTe, o-GeTe, r-GeSb$_2$Te$_4$ and s-GeSb$_2$Te$_4$

The atomic structures of pure r-GeTe, o-GeTe, r-GeSb$_2$Te$_4$ and s-GeSb$_2$Te$_4$, are shown in Fig. 1. The calculated lattice constant of the primitive cell of r-GeTe is a=b=c=4.27 Å and α=β=γ=59.50º, in close agreement with the experimental values [36]. This is a distorted simple cubic structure in which atoms move off-centre to form three short bonds and three long bonds. As in r-GeTe, the calculated r-GeSb$_2$Te$_4$ also undergoes such Peierls distortion in which the six bonds around each Ge site become three primary stronger bonds and three secondary weak bonds. This bonding is called resonant in which the primary bonds resonate between the primary and secondary positions [ref in Huang]. In the calculated o-GeTe, the Ge and Te are still three-fold bonded as in r-GeTe but the secondary resonant interlayer bonding is lost. In s-GeSb$_2$Te$_4$, the Ge atoms occupy tetrahedral positions, whereas the octahedral arrangement remains for the Te and Sb atoms. The sX local density of states (LDOSs) of each compound are shown in Fig. 2. The sX band gaps of r-GeTe and o-GeTe are 0.33 eV and 0.80 eV, respectively. Different experimental lowest band gap energy values have been reported for r-GeTe. Park et al [37] found this value to be 0.61 eV by linear extrapolation of the adsorption coefficient. Previous tunneling spectroscopy [38] and concentration dependence of susceptibility mass study [39] suggested a smaller energy gap around 0.1-0.2 eV. The lowest band gap energy values of 0.8 eV for the amorphous GeTe by the extrapolation method [37] and tunneling spectroscopy [40] agreed with each other. Note that o-GeTe is proposed as a lower symmetry crystalline approximation to the real amorphous GeTe and its sX band gap coincides with the experimental value. Theoretically, the calculated band gap values of r-GeTe varied from 0.367 eV to 0.66 eV based on the LDA functional [23,37,41], quantum mote carlo method gave even larger value of 0.7 eV [41]. Previous relativistic band structure calculations showed smaller band gap values, assuming the GeTe in the fcc structure [42,43]. For r-GeSb$_2$Te$_4$ and s-GeSb$_2$Te$_4$, the sX band gaps are 0.63 eV and 0.67 eV, respectively. The experimental extrapolation gave 0.49 eV and 0.71 eV for the crystalline and amorphous phases, respectively [37] and the LDA band gap for the crystalline phase is 0.55 eV [37]. The LDOSs (not shown) for both GeTe and GeSb$_2$Te$_4$ system show that the p states of the constituent atoms lie well above the s states, indicating weak s and p hybridization. This is an important requirement of the formation of resonance bonding [44]. The upper part of the valence band comprises of p states, roughly 2/3 from Te and 1/3 from Ge for GeTe, 2/9 from Ge, 1/3 from Sb and 4/9 from Te for GeSb$_1$Te$_4$ from the integral DOS (not shown), in line with the stoichiometry.

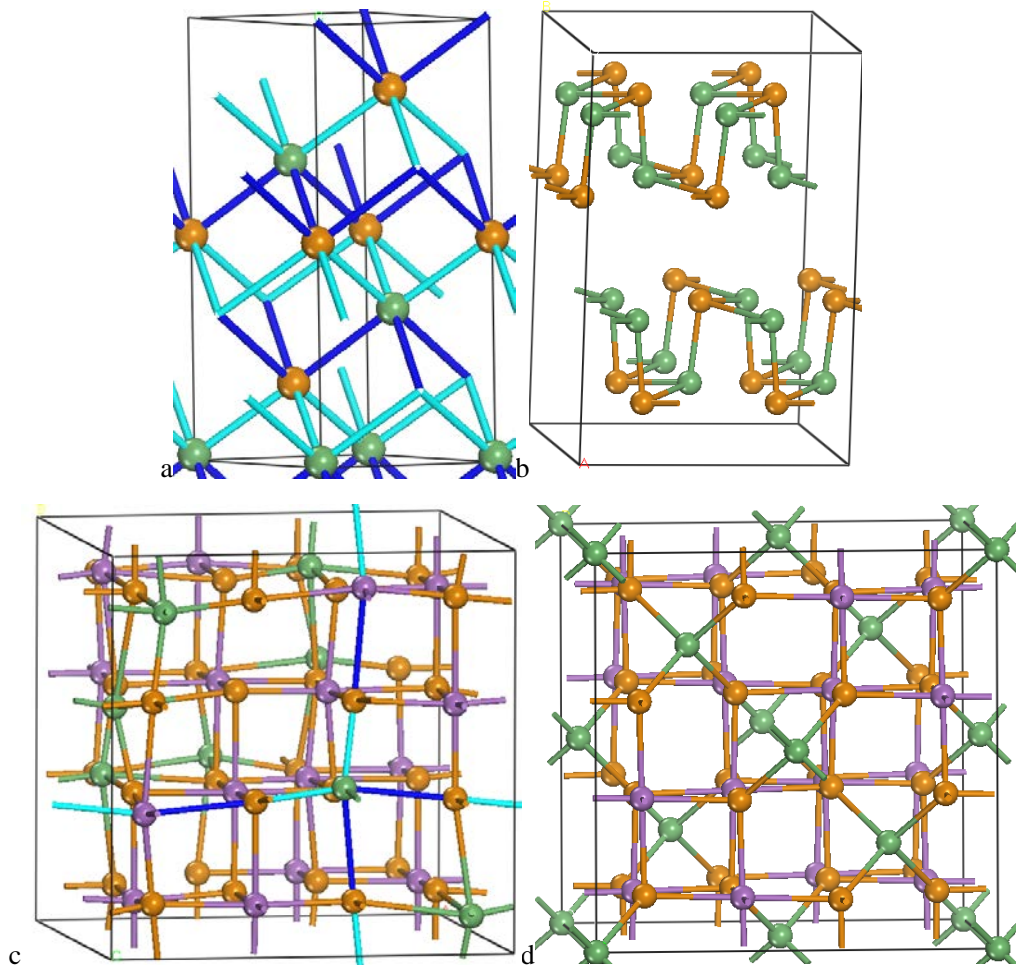

Fig. 1 Atomic structure of (a) r-GeTe, (b) o-GeTe, (c) r-GeSb$_2$Te$_4$, (d) s-GeSb$_2$Te$_4$. The long and short bonds in the resonance bond pair are denoted by light and dark blue, respectively, in (a) and (c).

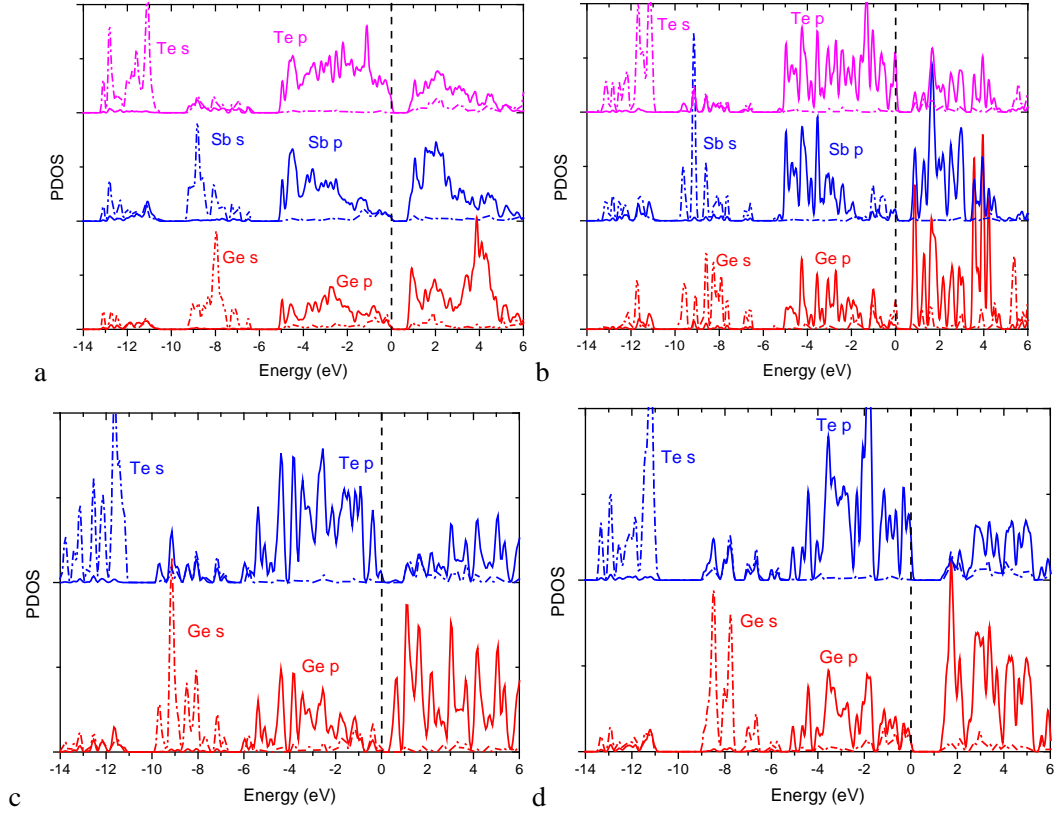

Fig. 2 LDOSs of constituent atoms in (a) r-GeSb$_2$Te$_4$, (b) s-GeSb$_2$Te$_4$, (c) r-GeTe and (d) o-GeTe. The solid and dash dot lines stand for angular momentum projected p and s states, respectively. Noting that the s and p states do not hybridize significantly.

It has been found that TM substitution at the Ge or Sb site is energetically favorable [14-19]. It can be viewed as the interplay between elemental TM atom and Ge or Sb vacancy. It is well known that excessive vacancies are presented in GST, which are speculated to be related to the p-type conductivity, switching mechanism, high crystallization speed, metal–insulator transitions in crystalline GST [45,46 and ref. therein], etc. It has been demonstrated by first principle calculation that Ge and Sb vacancies are most easily formed while the formation energy of Te vacancy is high [41,47]. Therefore, we only consider TM atom on Ge or Sb site in this work. We show in Fig. 3 the LDOSs of single neutral Ge and Sb vacancy, respectively, in r-GeSb$_2$Te$_4$ (r-GeTe similarly). Edwards et al [41] found that the geometry relaxation around the single vacancy in r-GeTe is small. We use the unrelaxed vacancy configurations here. We find that Ge vacancy introduces delocalized holes on the top of the valence band. According to the above component analysis of the valence band, Ge vacancy releases two electrons and one p state. The empty state where the holes stay comes from a pulled down host conduction band state and the total number of the valence band states is restored. Sb vacancy releases three p electrons and two p states. Similar to Ge vacancy, two host conduction p states are pulled down and the total number of the valence band states is restored. Therefore, there are three delocalized holes on the top of the valence band. These results are in agreement with previous LDA calculations [41]. We now consider the point defects in s-GeSb$_2$Te$_4$. For Ge vacancy, there is also a host conduction p state pulled down, as in r-GeSb$_2$Te$_4$. However, it forms an empty shallow acceptor state localized on the neighboring Te atoms rather than delocalized band state. For Sb vacancy, there is a shallow empty acceptor state on the top of the valence band below which another localized half-filled dangling bond state exists. These two states are mainly localized on the neighboring Te atoms which also come from two pulled down host conduction p states. There are equal number of interacting orbitals, regardless of the localization property, in the vicinity of the VBM for both defective r-GeSb$_2$Te$_4$ and s-GeSb$_2$Te$_4$ to interact with the substitutional TM atom.

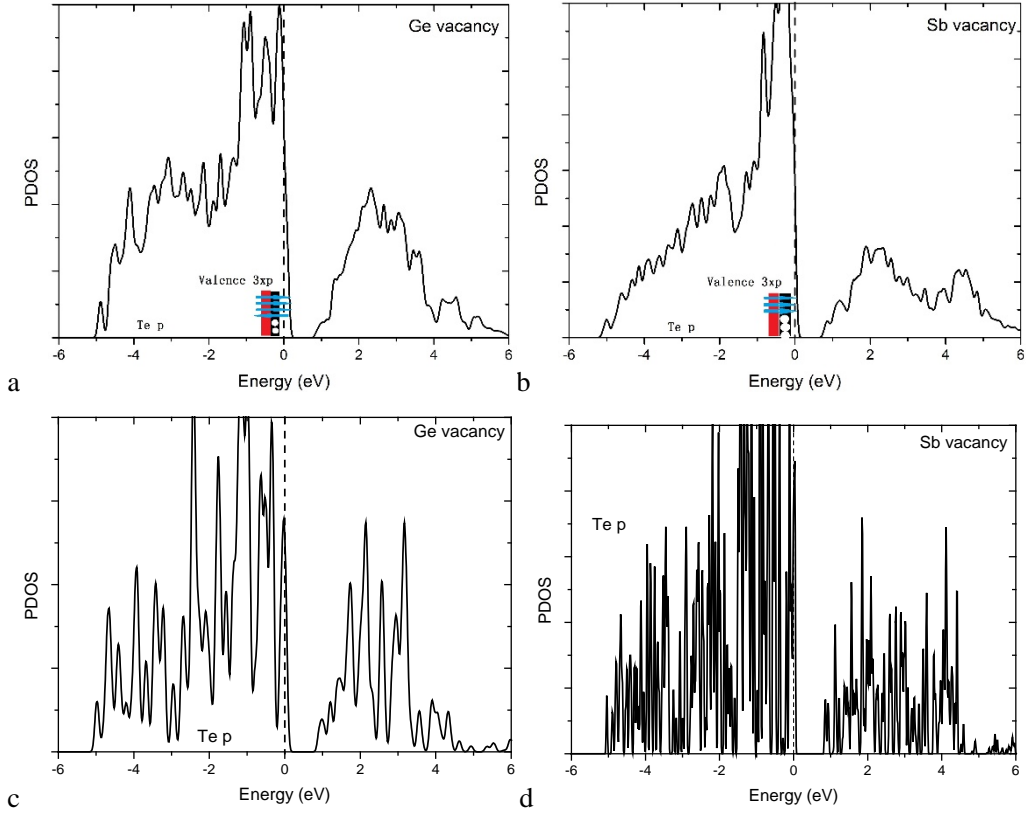

Fig. 3 LDOSs on atoms near the (a) Ge vacancy, (b) Sb vacancy in r-GeSb$_2$Te$_4$. The interacting orbitals relevant to the interaction with TM atom are illustrated. LDOSs on atoms near the (c) Ge vacancy, (d) Sb vacancy in s-GeSb$_2$Te$_4$. Noting that there are equal number of interacting orbitals in the vicinity of the VBM for each vacancy in both r-GeSb$_2$Te$_4$ and s-GeSb$_2$Te$_4$.

TM doped r-GeTe and r-GeSb$_2$Te$_4$

We substitute one Ge or Sb atom by Cr, Mn, Fe, Co and Ni atom, respectively, for both GeTe and GeSb$_2$Te$_4$ systems. These substitutional point defects are constraint in the neutral charge state. These substitution configurations are relaxed by the GGA functional and the electronic structures and magnetic properties of the relaxed structures are calculated by the sX functional. Fig. 4 shows the local magnetic moment on each substitutional TM atom and the total magnetic moment of the corresponding supercell for GeTe and GeSb$_2$Te$_4$, respectively. We first focus on the crystalline phases, namely r-GeTe and r-GeSb$_2$Te$_4$. A common feature for different substitutional sites of these two material systems is that the curve as a function of the TM atom with increasing nuclear charge number shows a cusp at Fe. We start from the discussion of Fe substitution.

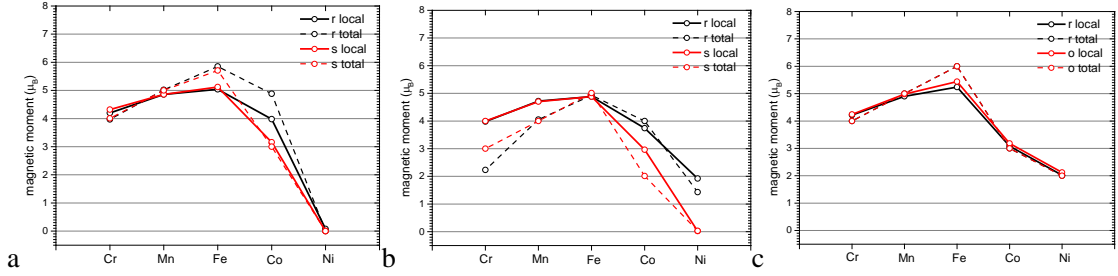

Fig. 4 The local and total magnetic moments of (a) $TM_{Ge}$ in $GeSb_2Te_4$, (b) $TM_{Sb}$ in $GeSb_2Te_4$, (c) $TM_{Ge}$ in GeTe

The LDOSs of $Fe_{Ge}$ in r-$GeSb_2Te_4$ are shown in Fig. 5(a). The majority d states are significantly lower than the p states of the neighboring Te and the minority d states are high in energy. This is in contrast to previous LDA results [14-19] where the 3d states are described as too shallow, resulting in a large hybridization with the host p states. The integral DOS shows that the majority Fe d states are fully occupied by five electrons and the minority Fe d states are empty. A simple molecular orbital description as sketched in Fig. 5(b) is useful to describe the interaction of Fe with neighboring atoms, as has also been done for Mn doped Ge complex [21]. As mentioned above, removing one Ge atom results in one fewer p state but the total number of the valence band states is restored, as labeled by '3xp' on the left of Fig. 5(b), by a pulled down host conduction p state. For the majority component, the Fe d orbitals are lower in energy than the host valence p orbitals. They interact and give rise to the majority states labeled as '5xd' and '3xp' (in red) in the middle of Fig. 5(b). For the minority component, the host valence p orbitals are lower in energy than the Fe d orbitals. They interact and give rise to two minority states labeled '3xp' and '5xd' (in black) in the middle of Fig. 5(b). The valence band maximum (VBM) is found to undergo negative exchange (or antiferromagnetic exchange) spin splitting (obtained from the eigenvalues at the sampling K points), this is because the '3xp' states at the VBM are sandwiched in between the exchange split majority and minority Fe d states and the p-d hybridization results in an exchange splitting of these states opposite in direction to that of the Fe atom [48]. Note that there is no band carrier (hole) in this case, which is required by the p-d Zener exchange induced spin splitting [49]. The Fe s orbitals are higher in energy. One of the differences between $GeSb_2Te_4$ and other octet semiconductors like Si, Ge, GaAs, etc. is that in $GeSb_2Te_4$ the host s and p orbitals do not hybridize and removing one host atom does not leave any s state for the Fe s electrons to fill in; they have to filled in the host p states. Up to now, twelve electrons are involved in the orbital interaction: two Fe s electrons, six Fe d electrons and four Te valence p electrons. However, five majority Fe d states and six host valence p states (three majority and three minority) below the Fermi level can only hold eleven electrons. One more host conduction p state, labeled as '1xp', has to be involved to hold the rest one electron. Instead of donating the rest electron into the host conduction band and leaving the system metallic (or half-metallic), from the LDOSs we see that the system is still insulating. In fact, '1xp' state undergoes significant positive exchange (ferromagnetic exchange) spin splitting: its majority component is pulled down to below the Fermi level which looks like a part of the host valence band; but its minority component is still in the host conduction band. The pulled down majority component of '1xp' state appears as a filled gap state near the host valence band edge, as seen from the LDOSs, which is mainly localized on the neighboring Te atoms of the Fe impurity. In typical DMSs like Mn doped GaAs, the interaction for the conduction band is driven by direct potential exchange between the host s band electrons and TM d electrons, which is always ferromagnetic [50]. In GST, however, the s and p bands do not hybridize significantly and from Fig. 2 we can see that the conduction band mainly comprises p states. The ferromagnetic exchange for the conduction band observed here needs further study. Clearly, the substitutional Fe is in the high spin state with five d electrons in the majority channel and zero d electron in the minority channel. The total magnetic moment of the complex is 6$\mu_B$. The local and total magnetic moments drawn from this interpretive model are in agreement with the calculated results.

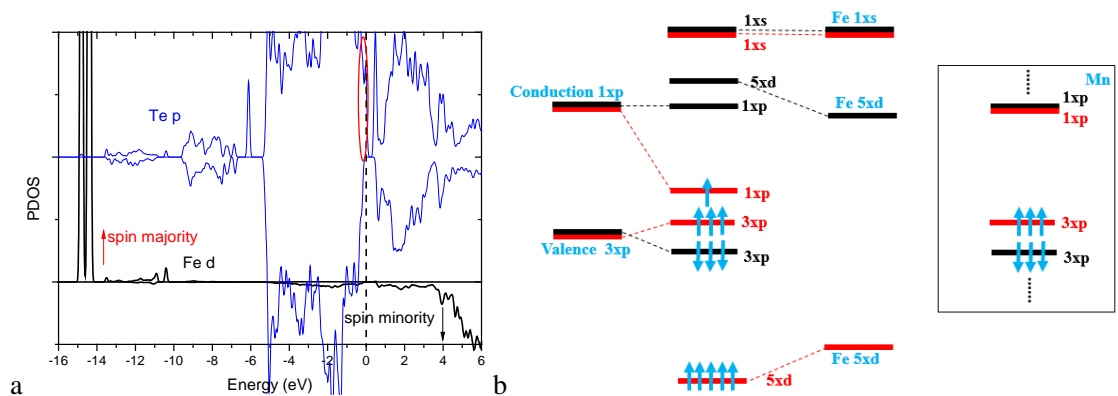

Fig. 5 (a) LDOSs of Fe$_{Ge}$ in r-GeSb$_2$Te$_4$, (b) schematic of the orbital interaction for Fe$_{Ge}$. The inset is for Mn$_{Ge}$.

The LDOSs of Fe$_{Sb}$ are shown in Fig. 6(a). The low lying majority Fe d states are fully filled while the high lying minority Fe d states are empty, the same as in Fe$_{Ge}$. The corresponding orbital interaction is sketched in Fig. 6(b). As mentioned above, similar to Ge vacancy, the missing valence p states from the Sb vacancy are compensated by two pulled down host conduction p states, we label them by '3xp' on the left of Fig. 6(b). Fe$_{Sb}$ has eleven electrons of relevance for the orbital interaction. Five majority Fe d states and six host valence p states (three majority and three minority) below the Fermi level hold all these electrons. The VBM is found to undergo negative exchange spin splitting. Therefore, the local magnetic moment on Fe and the total magnetic moments are both 5$\mu_B$, in agreement with the calculated results. A filled gap state near the host valence band edge is also found, which is mainly localized on the neighboring Te atoms of the Fe impurity.

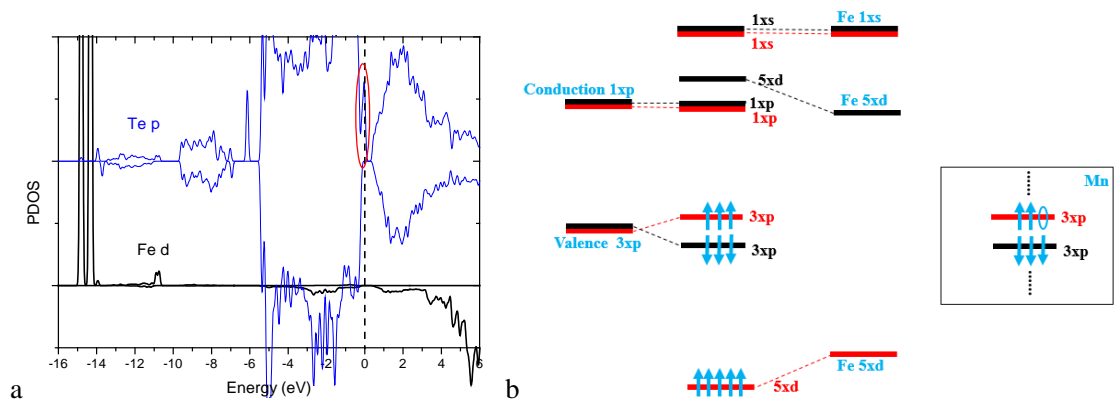

Fig. 6 (a) LDOSs of Fe$_{Sb}$ in r-GeSb$_2$Te$_4$, (b) schematic of the orbital interaction for Fe$_{Sb}$. The inset is for Mn$_{Sb}$.

The above discussion is based on the hybrid functional calculated electronic structures. Using

LDA/GGA calculated electronic structures, however, could give totally different magnetic properties and orbital interaction description, as have been done previously [14-19]. We briefly summarize the LDA results and the interpretation therein. As mentioned above, LDA tends to give the TM 3d states too shallow, too low the minority states or too high the majority states. Therefore, the TM 3d states turns out to hybridize with the host p states significantly. In such a case, a different orbital interaction description is needed as shown in Fig. 8. Taking the $Fe_{Ge}$ as an example, the minority states labeled as '5xd' in the middle of Fig. 8(a) are sufficiently low in energy to accommodate the rest one electron. There is no need to pull down another host p state from the conduction band to take the part. As a result, the local magnetic moment of Fe and the total magnetic moment are both $4\mu_B$. For $Fe_{Sb}$, it is analogous to $Fe_{Ge}$ except that the Fe minority d states are empty in this case. Therefore, the local and total magnetic moments are both $5\mu_B$. This orbital interaction description based on the LDA electronic structures is further simplified to an ionic model [14-19] that the Fe donates two and three electrons upon substitution of Ge and Sb, respectively, in order to form bonds with Te. The resulting electronic configurations of $Fe^{2+}$ and $Fe^{3+}$ ion are $3d^6$ and $3d^5$, respectively. Therefore, magnetic moments (total) of $4\mu_B$ and $5\mu_B$, respectively, are expected. However, the hybrid functionals like the sX split the majority and minority d states which are no longer too shallow. Such a LDA orbital interaction description is no longer appropriate that the local magnetic moments are different despite of the same total magnetic moments. The orbital interaction description in Fig. 8 is nevertheless helpful when the minority d states are sufficient low in energy, as the case of Co and Ni substitutions.

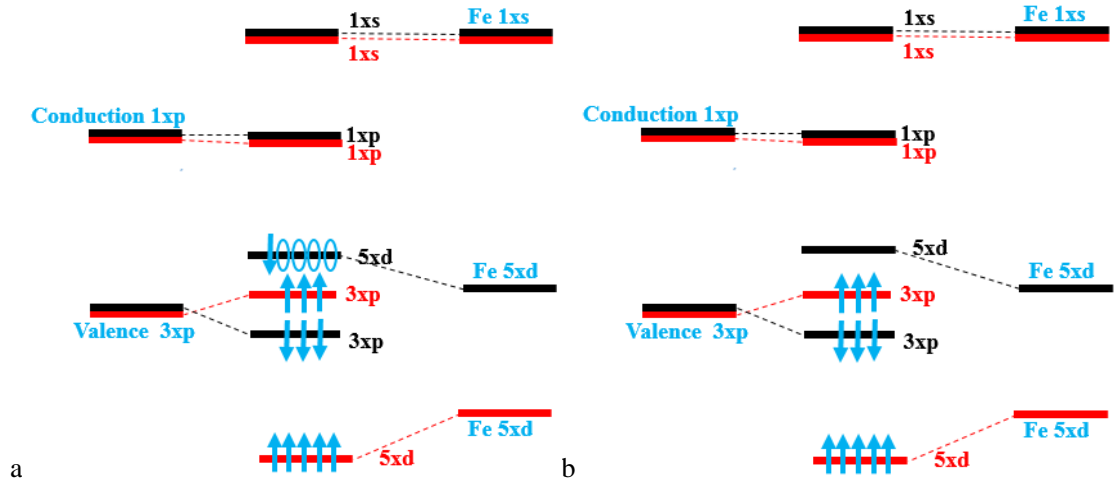

Fig. 8 Schematic of the orbital interaction for (a) $Fe_{Ge}$ and (b) $Fe_{Sb}$ in r-$GeSb_2Te_4$ based on previous LDA calculation. Note that the minority Fe d states are calculated to be too low that part of them are below the Fermi level.

Fig. 9(a) shows the LDOSs of $Co_{Ge}$ in r-$GeSb_2Te_4$. The Co majority d states are also deep blow the Fermi level and are fully occupied. The minority d states, on the other hand, are significantly lower in energy compared with that in $Fe_{Ge}$. Part of the minority d states are below the Fermi level and integral DOS shows that it occupies one electron. Co has one more electron than Fe and thus thirteen electrons are relevant for the interaction. The orbital interaction is sketched in Fig. 9(b). This configuration leads to local Co magnetic moment of $4\mu_B$ and total magnetic moment of $5\mu_B$. As in $Fe_{Ge}$, a filled gap state near the host valence band edge is found which is mainly localized on the neighboring Te atoms of the Co impurity. On the other hand, $Co_{Ge}$ in r-GeTe results in both the local and total magnetic moments of 3μ$_B$. This is because the majority '1xp' state is above the Fermi energy (orbital interaction not shown) and the minority Co d states are filled with two electrons as confirmed by the integral DOS (LDOSs not shown). In the case of Co$_{Sb}$, there are twelve relevant electrons. The LDOSs are shown in Fig. 9(c). The integral DOS also shows fully occupied Co majority d states and partially filled Co minority states by one electron. The orbital interaction is sketched in Fig. 9(d). This results in both local and total magnetic moments of 4μ$_B$, in agreement with the calculated results.

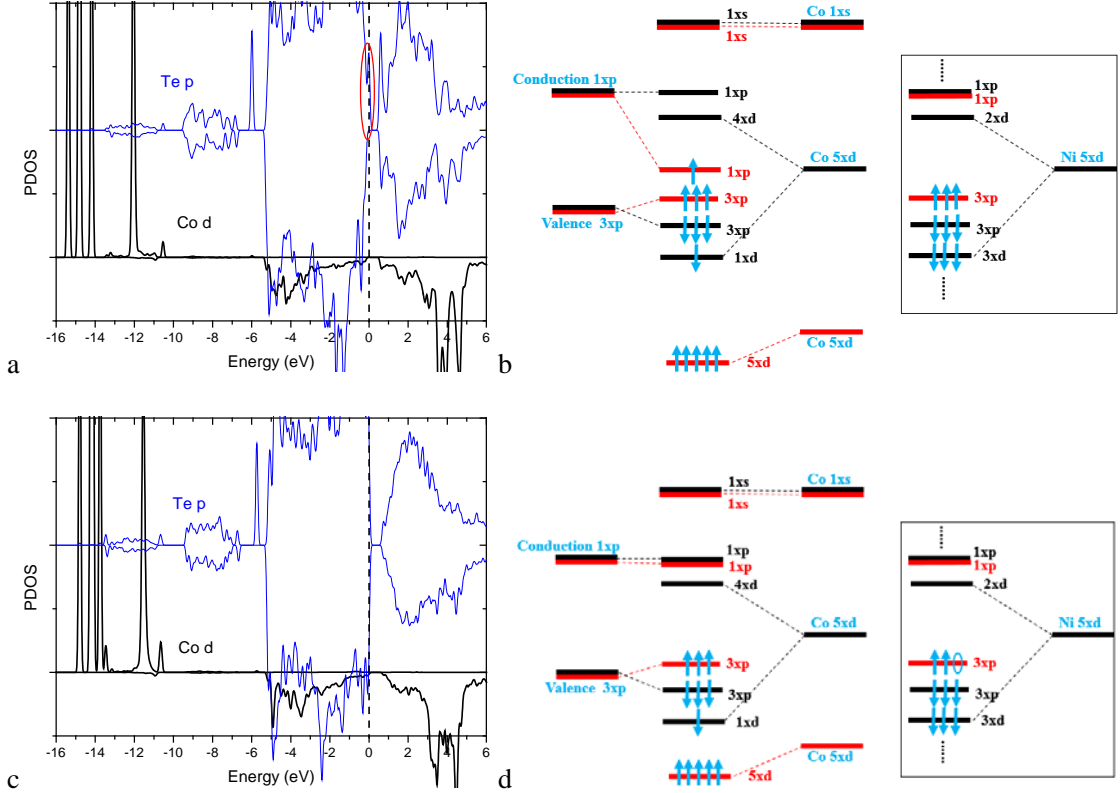

Fig. 9 (a) LDOSs of Co$_{Ge}$ in r-GeSb$_2$Te$_4$, (b) schematic of the orbital interaction for Co$_{Ge}$. The inset is for Ni$_{Ge}$ in r-GeTe. (c) LDOSs of Co$_{Sb}$ in r-GeSb$_2$Ge$_4$, (b) schematic of the orbital interaction for Co$_{Sb}$. The inset is for Ni$_{Sb}$.

For Ni$_{Ge}$ in r-GeSb$_2$Te$_4$, there are in total fourteen electrons of relevance. It is found that the majority spin states and the minority spin states are almost non-split (LDOSs not shown). Part of the Ni d states are lower than Fermi level. Integral DOS shows that both the majority and minority Ni d states are filled with four electrons. The host valence p states are both fully occupied. Ni$_{Ge}$ in r-GeSb$_2$Te$_4$ is nonmagnetic, namely both the local and total magnetic moments are 0μ$_B$. In r-GeTe, on the other hand, Ni$_{Ge}$ is magnetic. Integral DOS (LDOSs in Fig. 10(a)) shows that the Ni majority d states are fully occupied by five electrons and the minority d states are occupied by three electrons. The orbital interaction is similar to that of Co$_{Ge}$, as shown in the inset of Fig. 9(b), except that the majority '1xp' state in this case is higher than the Fermi level and the minority Ni d states hold three electrons. Therefore, both the local and total magnetic moments are 2μ$_B$. Explicit calculation of the local exchange by the sX functional to find out the energy difference of the ferromagnetic and nonmagnetic states is useful and is the subject of our future work. Fig. 10(b) shows the LDOSs of Ni$_{Sb}$ in r-GeSb$_2$Te$_4$. Integral DOS shows that the minority Ni d states are filled by three electrons.

An empty gap state near the host valence band edge is found which is mainly localized on the neighboring Te atoms of the Ni impurity. The orbital interaction is shown in the inset of Fig. 9(d). The local and total magnetic moments are $2\mu_B$ amd $0\mu_B$, respectively, close to the calculated values (Fig. 4(b) shows the value of the total magnetic moment between $1\mu_B$ and $2\mu_B$, but $2\mu_B$ is ruled out since the remaining five p electrons simply must not have $0\mu_B$ total moment).

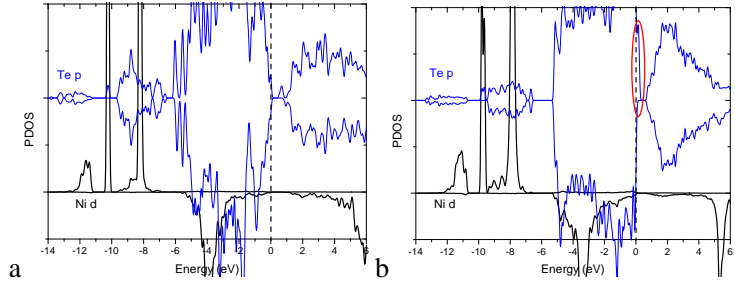

Fig. 10 LDOSs of (a) $Ni_{Ge}$ in r-GeTe, (b) $Ni_{Sb}$ in r-GeSb$_2$Te$_4$.

We now move to the other side of Fe to consider Mn and Cr. Let's first discuss Mn substitution. Mn has one electron fewer than Fe. The LDOSs of $Mn_{Ge}$ in r-GeSb$_2$Te$_4$ are shown in Fig. 11(a). We see that the majority d states are low and fully occupied, whereas the minority states are high and empty. This is also true in the case of Fe substitution case and similar orbital interaction description applies, as shown in the inset of Fig. 5(b). This results in both local and total magnetic moments of $5\mu_B$. The existence of an empty gap state near the host VBM cannot be accounted for by the simply orbital interaction description as it is the consequence of atomic structure relaxation near the impurity and the interaction among a larger set of orbitals. It is similar for $Mn_{Ge}$ in r-GeTe. For $Mn_{Sb}$, the LDOSs are shown in Fig. 11(b) and the corresponding orbital interaction is shown in the inset of Fig. 6(b). There are ten electrons of relevance so there is a hole at the majority '3xp' states which appears as an empty gap state near the VBM and is mainly localized on the neighboring Te atoms of the Mn impurity (see LDOSs). This results in local and total magnetic moments of $5\mu_B$ and $4\mu_B$, respectively.

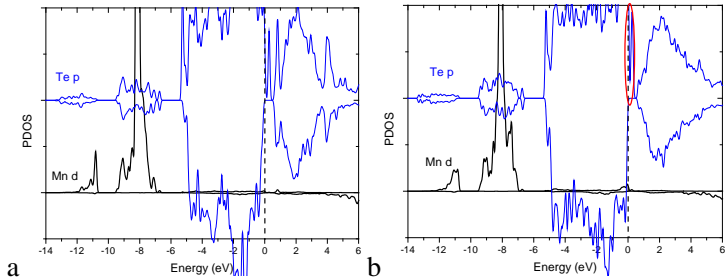

Fig. 11 LDOSs of (a) $Mn_{Ge}$ in r-GeSb$_2$Te$_4$, (b) $Mn_{Sb}$ in r-GeSb$_2$Te$_4$.

For $Cr_{Ge}$ in r-GeSb$_2$Sb$_4$, there are ten relevant electrons. The LDOSs in Fig. 12(a) show that the majority Co d states are high and part of them are above the Fermi level. The integral DOS shows that the majority Cr d states are filled with only four electrons and the minority Cr d states are empty. The orbital interaction is sketched in Fig. 12(b). The host valence '3xp' states are fully occupied with negative exchange spin splitting. The local and total magnetic moments are both $4\mu_B$. We also find a filled gap state near the host valence band edge. Unlike above mentioned gap states in other $TM_{Ge}$ complexes, however, the gap state in $Cr_{Ge}$ shows not only Te p character but also Cr d character. The decreasing of d character of the gap state from early 3d transition metal Cr impurity to heavier ones like Mn impurity in GaAs was also been found by LDA calculation [48]. $Cr_{Ge}$ in r-GeTe is similar. For $Cr_{Sb}$, there is one fewer electron. The integral DOS shows four electrons occupying the majority Cr d states (LDOSs in Fig. 12(c)). The orbital interaction is sketched in Fig. 12(d). In this case, there is a hole in the majority '3xp' states. The local and total magnetic moments are $4\mu_B$ and $3\mu_B$, respectively (Fig. 4(b) shows the value of the total magnetic moment between $2\mu_B$ and $3\mu_B$. $2\mu_B$ is ruled out since the remaining five p electrons must not have $-2\mu_B$ total moment).

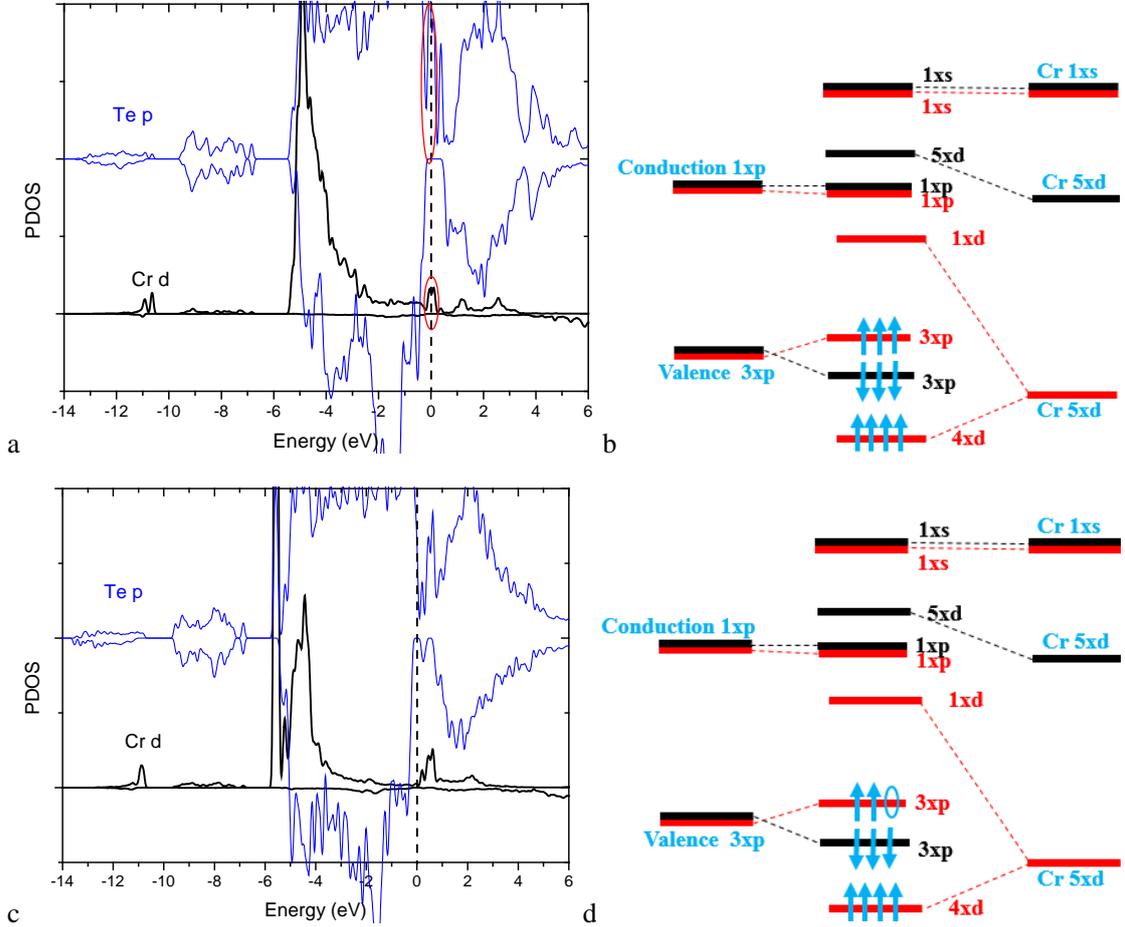

Fig. 12 (a) LDOSs of $Cr_{Ge}$ in r-$GeSb_2Te_4$, (b) schematic of the orbital interaction for $Cr_{Ge}$, (c) LDOSs of $Cr_{Sb}$ in r-$GeSb_2Ge_4$, (d) schematic of the orbital interaction for $Cr_{Sb}$.

From the above calculated LDOSs and the orbital interaction description of each TM doped r-$GeSb_2Te_4$/r-GeTe, we see that from early transition metal Cr to heavier Ni, the majority 3d states are gradually populated until fully occupied and then the minority 3d states begin to be filled. This trend generally agrees with that predicted by the LDA [14-19]. Note that in ref. [14-19] some results are presented for multiple interacting TM impurities but the corresponding magnetic properties are found to be qualitatively similar to those of the single impurity models [15]. In contrast to the LDA results, the sX results do not show significant DOS on Fe minority 3d states at the Fermi level, which are in fact empty. Instead, the sX results reveal large ferromagnetic exchange for the conduction band in the case of $Fe_{Ge}$, where the majority component of one conduction p state is pulled down to below the Fermi level and very close to the valence band edge, while its minority component is still in the conduction band. The readiness for orbital interaction of the conduction p states is also demonstrated in the LDOSs of the native vacancy (Fig. 3(a,b)). Due to the shift of Fe

3d states away from the Fermi level, the dominant TM-TM double exchange effect in stabilizing the ferromagnetism as speculated from the LDA calculation of the multiple interacting impurities model needs re-examining by the hybrid functional.

## TM doped o-GeTe and s-GeSb$_2$Te$_4$

In this part, we will briefly summarize the results of TM doped o-GeTe and s-GeSb$_2$Te$_4$ and compare the differences with the corresponding TM doped crystalline phases. For TM$_{Ge}$ in GeSb$_2$Te$_4$, magnetic contrast occurs for Co substitution. Co$_{Ge}$ in r-GeSb$_2$Te$_4$ shows local and total magnetic moments of 4$\mu_B$ and 5$\mu_B$, respectively, whereas in s-GeSb$_2$Te$_4$ it reduces to 3$\mu_B$ for both. The LDOSs of Co$_{Ge}$ in s-GeSb$_2$Te$_4$ are shown in Fig. 13(a). The majority Co d states are low lying and fully occupied while part of the minority Co d states are still above the Fermi level. The integral DOS shows that the minority Co d states are filled by two electrons instead of one as in r-GeSb$_2$Te$_4$. The orbital interaction is shown in Fig. 13(b). The resulting local and total magnetic moments are both 3$\mu_B$, in agreement with the calculated values. For other TM$_{Ge}$ in r-GeSb$_2$Te$_4$ and s-GeSb$_2$Te$_4$, the calculated magnetic moments are close which can also be interpreted by analogous orbital interaction model. TM$_{Ge}$ in o-GeTe and r-GeTe do not show any magnetic contrast.

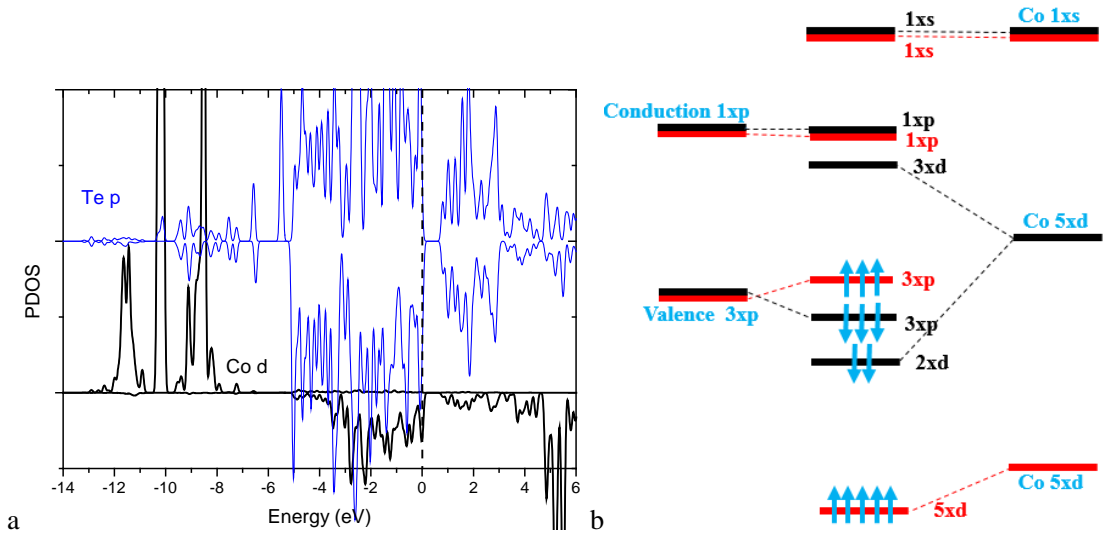

Fig. 13 (a) LDOSs of Co$_{Ge}$ in s-GeSb$_2$Te$_4$, (b) schematic of the orbital interaction for Co$_{Ge}$.

For TM$_{Sb}$ in GeSb$_2$Te$_4$, Co substitution also shows lower magnetic moment in the lower symmetry crystalline model s-GeSb$_2$Te$_4$. The integral DOS (LDOSs in Fig. 14(a)) of Co$_{Sb}$ shows that the majority Co d states are fully occupied and the minority Co d states are filled with two electrons instead of one as in r-GeSb$_2$Te$_4$. The orbital interaction is shown in Fig. 14(b). The resulting local and total magnetic moments are 3$\mu_B$ and 2$\mu_B$, respectively, in agreement with the calculated values. For Ni$_{Sb}$ in s-GeSb$_2$Te$_4$, it also shows lower magnetic moment (nonmagnetic in fact) than in r-GeSb$_2$Te$_4$. There is no spin splitting (LDOSs not shown). It has to be pointed out that by drawing the analogy between TM doped r-GeSb$_2$Te$_4$ and s-GeSb$_2$Te$_4$, the Ge/Sb vacancy induced dangling bond states in s-GeSb$_2$Te$_4$ are considered as part of the interacting valence p orbitals, without explicit treatment of their localization property. For other TM$_{Sb}$, the calculated magnetic moments in r-GeSb$_2$Te$_4$ and s-GeSb$_2$Te$_4$ are close which can also be interpreted by analogous orbital interaction model.

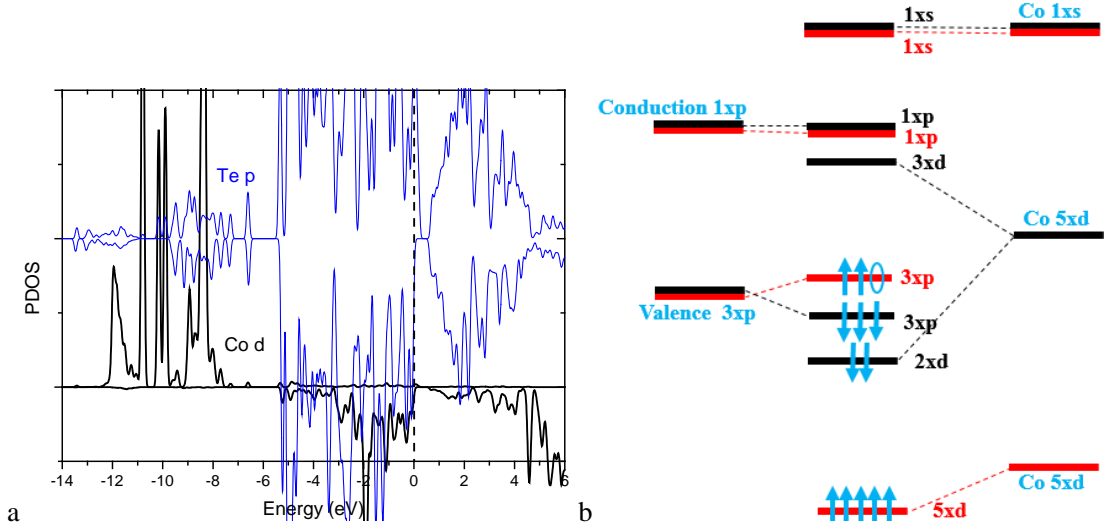

Fig. 14 (a) LDOSs of $Co_{Sb}$ in s-$GeSb_2Te_4$, (b) schematic of the orbital interaction for $Co_{Sb}$.

Experimentally, magnetic contrast occurs for ~7% [7] and 2% [8] concentration of Fe impurities in GST and GeTe, respectively. LDA calculation on TM doped crystalline and melt-and-quench MD generated amorphous GST and GeTe showed magnetic contrast [15,18], although the spin state is still under debate [14,19]. There are several possible reasons for the lack of magnetic contrast in our calculation. It is known that the p-d exchange mechanism stabilizes the ferromagnetism and is enhanced by the presence of excessive hole concentration from the Ge/Sb vacancies. Experimentally, the amorphous phase has less hole concentration than the crystalline phase so that the p-d exchange is less relevant. Our models do not take the hole concentration into account. Besides, our models are for single impurity where TM-TM interaction is not included. It is likely that TM-TM interaction in different phases has non-negligible effect on the magnetic contrast. In addition, our amorphous models are approximated by lower symmetry crystalline phases, which lack more complex structural patterns [29,30] of the real amorphous phases. It is possible that such patterns are important for the occurrence of magnetic contrast.

Conclusion

In this work, we carry out sX hybrid functional study of 3d TM doped $GeSb_1Te_4$ and GeTe. By curing the problem of LDA that the 3d states are predicted to be too shallow, sX shows different magnetic properties than those by LDA. The general trend is that from early transition metal Cr to heavier Ni, the majority 3d states are gradually populated until fully occupied and then the minority 3d states begin to be filled. The host VBM is found to undergo negative exchange spin splitting and is explainable by the p-d exchange. We also find that the host conduction p states are active for orbital interaction and they can undergo large positive exchange spin splitting upon interaction. The interaction between TM atom and the host GST is understood by simple orbital interaction description, which gives magnetic properties close to the calculated values. In comparing the magnetic properties between the TM doped crystalline and amorphous phases, we do not find obvious magnetic contrast for most TM impurities, including Fe. The discrepancy between the experiment and our calculation suggests the possibilities of hole concentration, TM-TM interaction, more complex characteristic structural patterns, etc. in determining the magnetic contrast.

This work was supported by high performance computing service of Tsinghua National Laboratory for Information Science and Technology.